\documentclass[journal,12pt,onecolumn,]{IEEEtran}
\usepackage{array}
\usepackage{color}
\usepackage{grffile}
\RequirePackage{afterpage}
\usepackage{epsfig}
\usepackage{here}
\usepackage{latexsym}
\usepackage{graphicx}
\usepackage{times}
\usepackage{program}
\usepackage{longtable,lscape}
\usepackage{subfigure}
\usepackage{rotating}
\usepackage{multirow}
\usepackage[linesnumbered,ruled]{algorithm2e}

\begin{document}


\title{Performance Evaluation of LTE-CommSense System for Discriminating the Presence of Multiple Objects in Outdoor Environment}

\author{\IEEEauthorblockN{Santu Sardar\IEEEauthorrefmark{1},
Amit K. Mishra\IEEEauthorrefmark{2},~\IEEEmembership{Senior Member,~IEEE} and
Mohammed Zafar Ali Khan\IEEEauthorrefmark{3}, ~\IEEEmembership{Senior Member,~IEEE}}
\IEEEauthorblockA{\IEEEauthorrefmark{1}ANURAG, Defence Research \& Development Organization, India}
\IEEEauthorblockA{\IEEEauthorrefmark{2}Department of Electrical Engineering, University of Cape Town, South Africa}
\IEEEauthorblockA{\IEEEauthorrefmark{3}Department of Electrical Engineering, IIT Hyderabad, India}
\thanks{This paper is a postprint of a paper submitted to and accepted for publication in  IEEE Transactions on Instrumentation and Measurement and is subject to  IEEE Transactions on Instrumentation and Measurement Copyright.}

}


\IEEEtitleabstractindextext{
\begin{abstract}
LTE-CommSense is a novel instrumentation scheme which analyzes channel affected reference signals of LTE downlink signal to obtain knowledge about the environmental change. This work presents the characterization of LTE-CommSense instrument to detect presence or absence of objects in outdoor environment. Additionally, we analyze its capability of detecting and distinguishing when multiple objects are present. For performance evaluation and characterization of this instrument, we derive object detection accuracy, FAR, FRR and resolution which we believe are the most important figures of merit in this case. As the operation of LTE-CommSense is to detect events instead of objects, we redefine the concept of resolution for LTE-CommSense. Two different proposals to represent the redefined resolution viz. Neyman Pearson principle based and Cramer Rao principle based resolution are presented here. All the performance metrics are derived using practical data captured using an SDR platform modeled as a LTE-CommSense receiver. We observe that, LTE-CommSense provides better performance in detecting presence or absence of objects at near range.
\end{abstract}
}

\maketitle

\section{Introduction}

Recently there has come an upsurge in activities in the field of using communication radiation for target detection and tracking using receiver only nodes (\cite{baker,tong,Huang2015LowPO,7997194,199427}). Using the same principle, our group has been working for the past few years on a novel remote sensing instrument which works by analyzing the channel equalization process of telecommunication systems. We call it communication based sensing or CommSense (\cite{akm_patent,bhatta,phd_mag}). We have shown that CommSense system can be used to distinguish different weather events, presence and absence of objects and changes in the immediate environment. In case of LTE-CommSense, it uses the Long-Term Evolution (LTE) downlink (DL) signal from the base station (BS) to the user equipment ($UE$) to extract the channel information and estimate environment change (\cite{phd_mag,7887698}). The reference pilot symbols transmitted at periodic intervals by the LTE base station or $eNodeB$ \cite{lte_book} for channel estimation and subsequent equalization purpose gets channel-affected while passing through the communication channel. From receiver channel estimation block LTE-CommSense instrument \cite{phd_mag} first obtains the channel properties in between $eNodeB$ and $UE$. The channel estimation block finds the channel impulse response (CIR) which represents the properties of the dynamic channel \cite{pheno}. It can be noted here that the CommSense type of sensing does not need any dedicated transmitter contrary to the proposals in (\cite{7831367,8360051,8349939,8326735,6582678}) and hence does not need to adhere to transmitter related regulations. The major novelty of CommSense lies in the fact that compared to existing systems (\cite{7835628,7905996,7472878}) on localization using WiFi or GSM, in our case we are using no transmitter and it is purely based on pilot carriers only.

In some of the previous reports from our work on CommSense we have shown some encouraging results using both GSM and LTE telecommunication standards (\cite{bhatta,phd_mag,7489757,africon}). However, we have not done a formal analysis of the performances of this newly proposed instrument. In the current paper, we apply LTE-CommSense for detection of presence of object in outdoor environment. After that, it was investigated that, whether it can detect and distinguish multiple objects or not. To perform that, we have identified the important figures of merit suitable for LTE-CommSense instrumentation scheme. The figures of merit which we believe to be the most important in this scenario are accuracy of detection, false acceptance rate (FAR), false rejection rate (FRR) and resolution. The concept of resolution is redefined as per the CommSense working philosophy \cite{4071623} of detecting events instead of objects. CommSense is targeted to detect environmental changes. Types of environmental change will depend on the exact type of application the CommSense instrument needs to detect. For the application in hand, the event may be defined as presence or absence of an object in the immediate outdoor environment. We proposed two methods to evaluate the resolution for LTE-CommSense viz. Neyman Pearson (NP) principle based resolution and Cramer Rao (CR) principle based resolution from two different vantage points. 

This paper is organized as follows. In Section $2$, the basic concept of LTE-CommSense is summarized. All the identified characterization parameters and need of redefinition and proposed resolution concept is explained in Section $3$. This section also provides two potential proposed definitions of resolution of LTE-CommSense. Section $4$ elaborates the experimental procedure to collect the data for detection and distinction of presence of multiple objects practically. The results, evaluated figures of merit, analysis of the extracted data and procedure of calculation of resolution according to the proposed definitions are detailed in Section $5$. Section $6$ provides the conclusion from these efforts.

\section{LTE-CommSense System}

\begin{figure}[!h]
\centering
\psfig{file=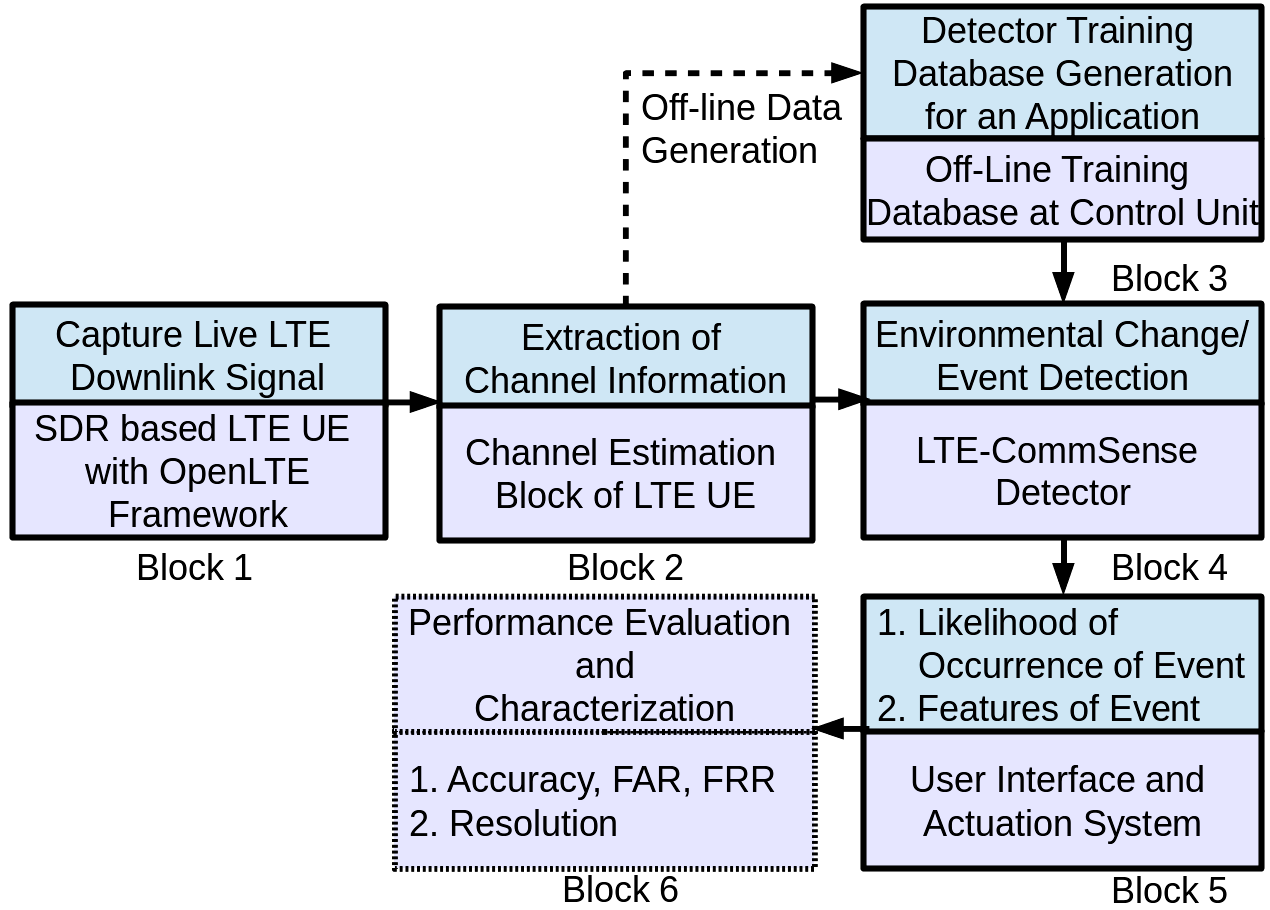,width=0.43\textwidth}
    \caption{LTE-CommSense instrumentation system block diagram using single $UE$.}
    \label{block}
\end{figure}

General block diagram and description of LTE-CommSense are elaborated in (\cite{africon,phd_mag}). Figure \ref{block} provides a part of the complete conceptual system described in (\cite{africon,phd_mag}) using single $UE$ which is implemented and utilized for this work. 

There are two parts of processing in the system: one is offline and the other one is online. Offline processing chain contains blocks 1, 2, 3 and 4 whereas online processing contains block 1, 2, 4 and 5 in consecutive order. Processing block 6 is particular to this work of characterization. A brief description of individual blocks are as follows.

\begin{itemize}
\item Block 1: This block first detects and then captures LTE DL signal from the $eNodeB$ using the USRP N200 software defined radio (SDR) platform, configured as an LTE $UE$ by the use of OpenLTE framework. This block is used both in offline and online modes. 
\item Block 2: In the LTE $UE$ receiver work-flow, the channel estimation block extracts the channel information from the channel affected cell specific reference symbols (cellRS). The objective of this block is to tap in to the channel estimation block of the $UE$ to capture the extracted channel information. This block is also used both in offline and online mode. 
\item Block 3: It is used only in offline mode. Based on the target application of object presence detection in outdoor environment and distinguishing presence of multiple objects, we first record LTE DL signal. Using the captured DL data, we have to perform LTE channel estimation and equalization to extract channel impulse response (CIR). Using the CIR values, we have to generate the database to train the LTE-CommSense detector. Data corresponding to multiple capture of application scenarios are used to generate the training database and maintained in the control unit. The control unit tunes LTE-CommSense for targeted application.
\item Block 4: This is the LTE-CommSense detector and used in both modes of operation. First, it works in offline mode and gets trained by the application specific database provided by the control unit in `Block 3'. After that in online mode, it accepts a channel information corresponding to a DL signal capture of current channel condition. The objective of the detector is to give decision of regarding the environmental change or event detection as per the targeted application. Support vector machine (SVM) based classifier is used for the training and testing in this block. It may be replaced with another classifier depending on the application at hand \cite{7882660}.
\item Block 5: This block conveys the decision of the detector to the user to take further action by the actuation system. This block works only in online mode. It contains the likelihood of the event occurrence. If the event has occurred, then this block may also provide relevant features of it. In the present case, first this block provides decision regarding presence or absence of an object. If present, it provides further decision of how many objects are present.
\item Block 6:  This block is for system performance analysis and characterization of LTE-CommSense for the targeted application.
\end{itemize}

\subsection{Feasibility of Using LTE Reference Signals as Bi-Static Radar Waveform}

\begin{figure}[h]
\centering  
\psfig{file=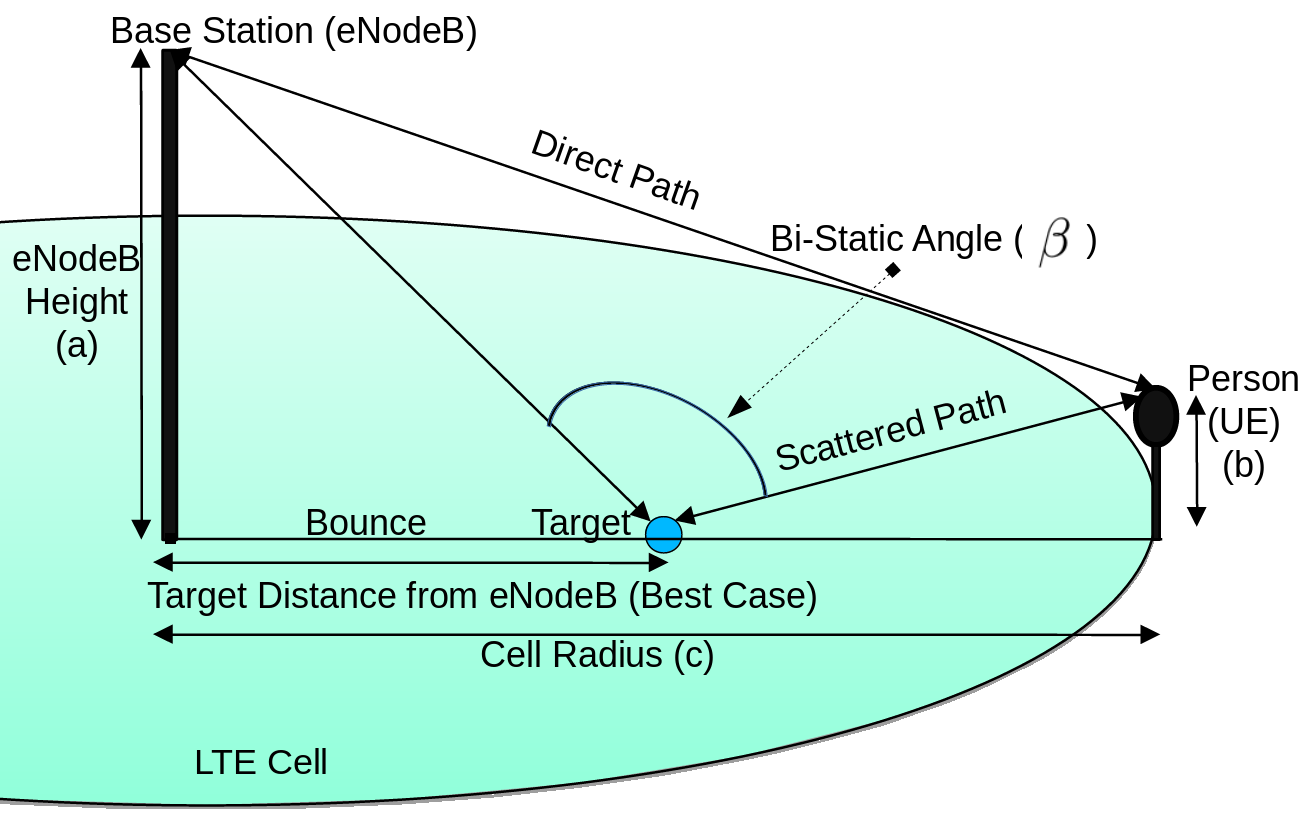,width=0.42\textwidth}  
\caption{Best Case Scenario for the calculation of Time Resolution Requirement to use LTE signal in Conventional Radar Framework}
\label{lte_radar_time}
\end{figure}

To detect environmental changes using LTE communication infrastructure, another alternative approach is to utilize the infrastructure as a conventional bi-static radar system. The static base station (BS) i.e. $eNodeB$ will work as the transmitter and $UE$ is the receiver. To utilize it as a radar system, we need a reliable precision clock to find out the time difference between the direct $DL$ signal path from $eNodeB$ to LTE $UE$ and the scattered path from the target. Figure \ref{lte_radar_time} shows the configuration of the base station, user equipment and the relative position of the target. According to the calculations described in Appendix, can be evaluated as shown in Table \ref{lte_time}. One alternative would be to utilize the clock signal sent by the LTE base stations. System information block (SIB) 16, available for LTE releases 11 onwards, contains Coordinated Universal Time (UTC). As per the standard, the precision of UTC time in this `$SIB-16$' is $10 ms$. We made some simple calculations to check if this is precise enough for us.  

It can be observed from Table \ref{lte_time} that, the timing precision requirements for the best case scenario of detecting a target (Figure \ref{lte_time}) are higher than the LTE specifications \cite{lte_spec_3}. Therefore, LTE reference signals may not be suitable to use as conventional radar. Hence we investigated alternate ways to extract usable information from the channel data. Without reliable clock we can not use the conventional definition of resolution. As shall be described later, we define alternate definitions of resolution to validate the performance of our system.

\begin{table}[!h]
\caption{Timing precision requirements for different types of LTE Cell and Base Stations to use LTE Downlink Reference Signals in conventional radar framework.}
\begin{center}
\begin{tabular}{|m{20mm}|m{15mm}|m{30mm}|}\hline
BS Type & Cell Type & Best Case Timing Precision Requirement \tabularnewline	\hline
Wide Area & Small &	79.9ps (3km radius)   	  \tabularnewline	\hline
Wide Area & Large &	39.9ps (6km radius)		  \tabularnewline	\hline
Home Area & Small &	0.48ns (0.5km radius) 		  \tabularnewline	\hline
Home Area & Large &	0.24ns (1km radius) \tabularnewline	\hline
\end{tabular}
\end{center}
\label{lte_time}
\end{table}

\section{Characterization Parameters}

\subsection{Detection Accuracy, FRR and FAR}
For this detector system, the accuracy (\cite{78300,4071623}) denotes the extent with which the system provides correct detection output i.e. the system's ability to distinguish between presence  and absence of the object in outdoor environment. 

Another two standard measures to indicate the identifying power of a detection methodology are FRR or false negative rate (Type I Error) and FAR or false positive rate (Type II Error) \cite{farfrr}. FAR is fraction of the falsely accepted patterns divided by the number of all impostor patterns. The fraction of the number of rejected client patterns divided by the total number of client patterns is called FRR. 

The frequency at which false acceptance errors are made can be denoted as FAR. If $N_{FA}$ is the number of false acceptance cases and $N_{IA}$ is the total number of impostor attempts, then FAR can be defined as: 

\begin{equation}
FAR = \biggr(\frac{N_{FA}}{N_{IA}}\biggr)
\end{equation}

Similarly, false rejection is the act of deciding that a category is an impostor while the category is actually genuine. The frequency at which false rejection are made is called FRR. If $N_{FR}$ is the total number of failse rejections, and $N_{EA}$ is the total number of legitimate classification attempts, then FRR can be given as: 

\begin{equation}
FRR = \biggr(\frac{N_{FR}}{N_{EA}}\biggr)
\end{equation}

\subsection{Resolution: Conventional Range Resolution and Proposed Redefinition}

The definition of resolution for any generic instrument according to the Measurement Systems Analysis Manual is: ``the resolution of an instrument is $\delta$ if there is an equal probability that the indicated value of any artifact, which differs from a reference standard by less than $\delta$, will be the same as the indicated value of the reference". Therefore generic instruments measure the ``value of any artifact" \textcolor{black}{\cite{Mishra16b}}. This general definition of resolution is applicable to any instrument. For a radar system we measure radar range resolution. For a bi-static radar system where the communication base-station antenna is considered to be the transmitter and the user equipment ($UE$) to be the receiver, we can find out the range resolution. According to general bi-static radar theory, range resolution of this LTE based commensal radar can be defined by:

\begin{equation}
\Delta R = \frac{c}{2*B*\cos(\frac{\beta}{2})}
\label{rad_range}
\end{equation}

Here, $\beta$ is called bi-static angle i.e. the angle the target object makes between the base station transmitter and the $UE$. The term $B$ denotes the signal bandwidth.

This conventional definition of `Resolution' is not applicable directly in this case because this instrument is not a conventional radar. Here we are not detecting objects; rather we are distinguishing events. In the analysis presented in this paper, the event is defined as: ``Detection of the presence of objects in outdoor environment". Additionally if presence of object is ascertained, then this instrument will estimate the number of objects present. Therefore, there is a basic difference between resolution of conventional object detector as detailed in Equation \ref{rad_range} and CommSense based environment change detector \cite{Mishra16b}. Instead of processing of radio frequency (RF) data, here the estimated channel impulse response (CIR) is used to distinguish events occurring in the channel. It should also be event dependent. Therefore `Resolution' has to be redefined here. As the objective is to distinguish events, the `Resolution' should be defined with respect to events. Consequently, the value of $\beta$ is not required here. Therefore direct comparison of the `Resolution' of LTE-CommSense with the traditional range resolution equation may not be relevant. The resolution of LTE-CommSense may be represented by making use of two different models:

\subsubsection{Neyman-Pearson Principle based Definition}

LTE-CommSense provides a likelihood or probability, $P_p$ which signifies that a certain `artifact of interest' is present. Let us represent the artifact of interest as $\theta$. For a given task there will always be cases of misclassification which will create events of false alarm giving a probability of false present, $P_{fp}$ . Then using Neyman Pearson \cite{Mishra16b} criteria, the goal of designing an LTE-CommSense detector is to maximize $P_p$ for a given $P_{fp}$.

Therefore, using Neyman Pearson principle, resolution $(\delta_{NP})$ can be defined as the maximum change in the artifact of interest, $\Delta \theta$, which for a given LTE-CommSense detector makes sure that the change in the probability of false present $\Delta P_{fp} \leq $ an arbitrary small number.
\\

\emph{\textbf{Definition 1:}}
The Neyman Pearson principle based resolution $(\delta_{NP})$ can be defined as the maximum change in the artifact of interest, $\Delta \theta$, for which the change in the probability of false present $\Delta P_{fp} \leq \epsilon$, $\epsilon$ being an arbitrary small number.
\\

\subsubsection{Cramer-Rao Bound based Definition} \label{crsec}

From a data-centric point of view for an `artifact of interest', $\theta$, we may view the LTE-CommSense detector system of LTE-CommSense to capture some signal $\zeta$. One of the major empowering concept of LTE-CommSense detector is the hypothesis that we can estimate $\theta$ from $\zeta$ even when the exact phenomenological link from $\theta$ to $\zeta$ is not well-modeled. Therefore, LTE-CommSense detector operation becomes necessarily a parameter-estimation operation. Now, in parameter estimation, one of the important figures of merit is the Cramer Rao lower bound (CRLB) which is the lowest variance \textcolor{black}{\cite{Mishra16b}} achievable in the estimate. According to the statement of the Cramer Rao theorem, if $\hat{\theta}$ is an unbiased estimator of $\theta$, then,

\begin{equation}
Var(\hat{\theta}) \geq \frac{1}{I(\theta)}
\label{crlb_eq}
\end{equation}

Here, $I(\theta)$ is a measure of average information and represented by Fisher information. Fisher information statistic can be expressed as:

\begin{equation}
I(\theta) = E\biggr[\biggr(\frac{\partial l(\zeta;\theta)}{\partial \theta}\biggr)^2\biggr]
\end{equation}

where $l(\zeta;\theta)$ is the natural logarithm of the likelihood function and $E$ represents the expectation operation. For ill-defined mapping like what we deal within LTE-CommSense detector, Fisher information can be found numerically. The lower bound shown in Equation \ref{crlb_eq} is achieved for an unbiased estimator of $\theta$. Therefore to obtain the Cramer Rao principle based resolution of LTE-CommSense, it is required to find the minimum variance of the estimator from measured data.
\\

\emph{\textbf{Definition 2:}}
The Cramer-Rao principle based resolution $(\delta_{CR})$ can be defined as the maximum change in the `artifact of interest', $\Delta \theta$, for which the change in the inverse of Fisher information $(\Delta I(\theta))$ or the variance of the estimate is less than or equal to an arbitrary small number $\epsilon$.

\section{Experimental Setup}\label{setupsection}

The objective of this set of practical experiments is to detect the presence or absence of objects in outdoor environment. If detected, this proposed system should further determine whether one or two targets are present. We derive the performance parameters and figures of merit along with the redefined resolution methodically for LTE-CommSense scenario from practical experiment for this case. 

\begin{figure*}
\centering
\subfigure[]{\psfig{file=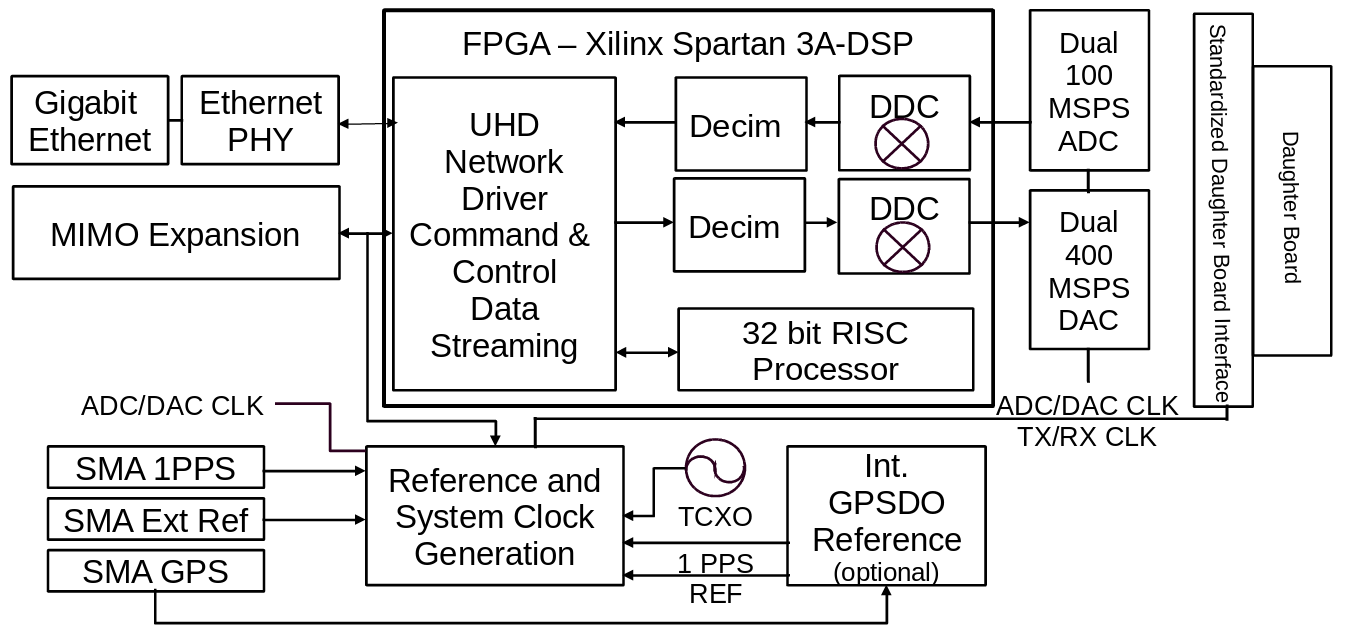,width=.49\textwidth}}  
\subfigure[]{\psfig{file=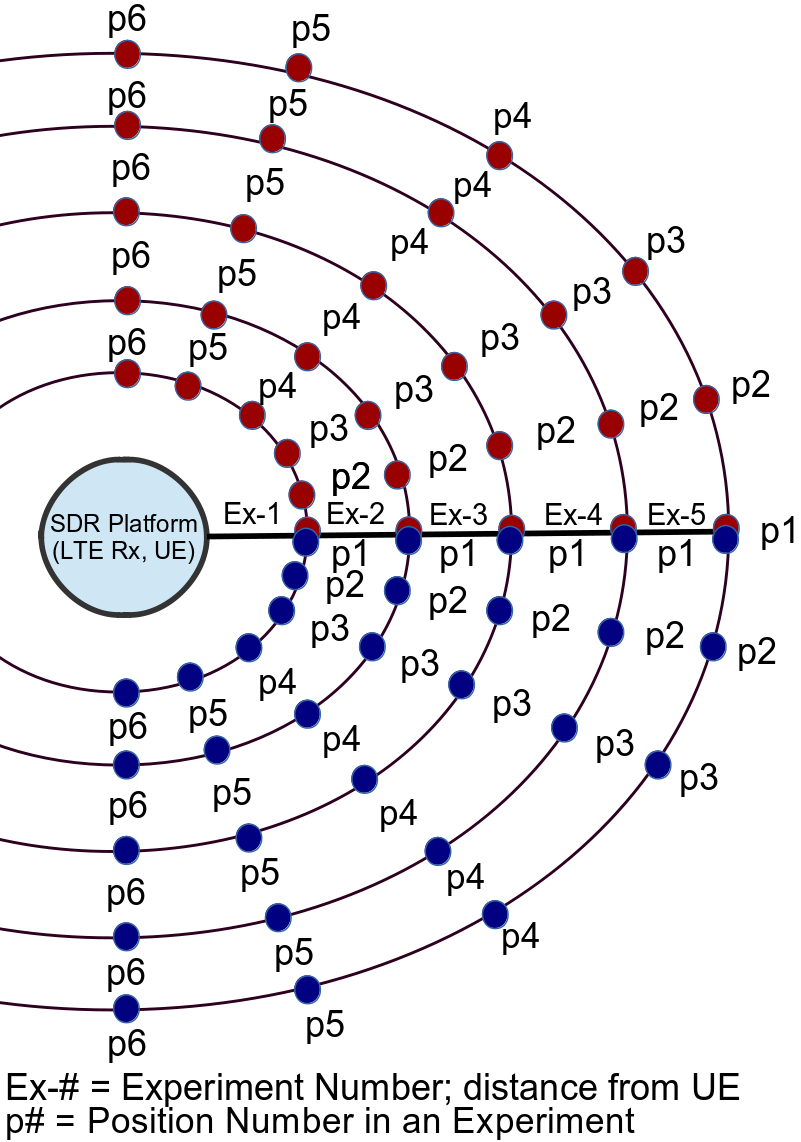,width=0.23\textwidth}} 
\caption{(a) USRP N200 Internal Block Diagram, (b) Experimental Setup to Calculate Resolution: The SDR based LTE $UE$ is located at the center of the concentric circles. These concentric circles represents different distances of the reflectors from the $UE$. For each specific distance of the reflectors from the $UE$, six different readings were taken ($p1,p2,p3,p4,p5,p6$). The small red and blue filled circles are the locations of the reflectors. For these six cases, LTE DL signal is captured by the $UE$.}
\label{usrpn200}
\end{figure*}

To obtain live LTE downlink signal, SDR based data capturing system was developed. USRP N200 SDR platform \cite{usrp} and RFX2400 RF daughter card \cite{rfx2400} along with VERT2450 antenna \cite{antenna} was used to work at LTE passband operating frequency range. A GNURadio based open source model of the 3GPP LTE receiver on OpenLTE \cite{openlte} was utilized to control the SDR platform. The internal block diagram of USRP SDR platform is shown in Figure \ref{usrpn200}(a). The SDR unit was connected to the host computer via Ethernet. The OpenLTE framework was implemented on the host computer. The commands for searching of live LTE DL signal, capturing detected LTE DL data on the host computer for a specified time duration were also input to the SDR platform via host PC using the OpenLTE framework. The LTE $UE$ receiver works at $Band-40$ (2300 MHz to 2400 MHz frequency band) in time division duplexing (TDD) topology. Figure \ref{usrpn200}(b) shows the SDR platform, relative placements of the objects for this experiment. Corner reflectors are considered in this practical experiment as reflectors. In this figure, the details of the ``{\it SDR platform (LTE Rx, $UE$)}" at the middle of the setup is shown in Figure \ref{usrpn200}(a). The details of the reflectors configuration and their placements are described as follows.

\subsection{Reflectors Configuration and Their Relative Placement}

In this work, the objective is to evaluate performance of the LTE-CommSense instrument in detecting presence or absence of objects. If presence is detected, it should also predict the number of objects. In this experiment, we have targeted to detect and distinguish two objects. Therefore we have used two corner reflectors as our targeted objects. These two reflectors are placed at different distances from each other and from the modeled LTE $UE$ to experiment about the distinguishing ability of LTE-CommSense. 

Corner reflectors have the advantage of having very high radar-cross-section (RCS) for relatively smaller size and the RCS is maintained over a wide range of incident angle. They are easy to make from metallic sheet such as Aluminum (Al). But care must be taken to ensure that the surfaces are joined at exactly $90^o$ degrees. Also, they should be robust enough to maintain good flatness. In this case, the reflectors are metallic sheets made of Aluminum (Al) with properly perpendicular faces. 

The size of the reflectors should be based on the wavelength of the LTE DL signal. The carrier frequency ($f$) of LTE is $~2.1$ GHz \cite{lte_spec_2}. Therefore, if the corresponding wavelength is denoted by $\lambda$, the relation between $f$ and $\lambda$ is: $ \lambda*f = C$. Here, $C$ is the speed of light in vacuum i.e. $C = 3*10^8$ m/s. Therefore,

\begin{equation}
\lambda = \frac{C}{f} = \frac{3*10^8}{2.1*10^9} = 0.14m
\end{equation}

We observe that, the wavelength of the LTE DL signal will be in the range of \textcolor{black}{$0.14m$}. The size of each face is made five times more to ensure sufficient amount of reflected signal is reflected and received by the modeled LTE $UE$ terminal on the SDR platform.

As shown in Figure \ref{usrpn200}(b), two different reflectors were placed at different distances from each other and different distances from $UE$. The environment selected was outdoor open field environment. Data were collected for one reflector and two reflectors cases. Five different distances of the reflectors from the $UE$ were considered viz. \textcolor{black}{$0.5m$, $2m$, $4m$, $7m$ and $10m$}. For each of the distances, intra-reflector distances of the two reflectors were varied along the circumference of the circle with the radius to be the distance from the $UE$ and center at the $UE$ position. Six different intra-reflector distances were considered as per the formula: 

\begin{equation}
S = x*\pi*D
\label{circ_dist}
\end{equation}

Here $D$ was the distance of the reflectors from the $UE$ and $x$ was a number which has assumed the values $0.00$, $0.10$, $0.25$, $0.50$, $0.75$ and $1.00$ respectively during the experiment. This distance was along the circumference of the circle which has $D$ as the radius. Correspondingly the straight line distance between the reflectors can be found using:

\begin{equation}
D_{intra} = 2*D* \sin \biggr(\frac{S}{2*D}\biggr)
\label{intra_abs}
\end{equation}

Here, $D_{intra}$ was the straight line distance between the two reflectors, $D$ was the radius of the circle i.e. distance of the reflectors from the $UE$ and $S$ was the distance of the reflectors along the circumference of the circle.

\section{Results and Analysis }

Using the experimental setup, the primary figures of merit viz. accuracy of detection of targeted event, FRR and FAR are evaluated first. Then, these are used to evaluate the resolution of LTE-CommSense instrument as per the proposed definitions.

\subsection{Figures of Merit}

Figure \ref{accu}(a) shows the variation of the average values of accuracy w.r.t. the absolute distance between the reflectors as well as their distances from the $UE$. The $X$ axis shows the distances (m) of the reflectors from the $UE$ whereas the $Y$ axis shows the absolute distance (m) between the reflectors. In this figure, the accuracy corresponding to every distance point is indicated. It can be observed that, for a particular distance of the reflectors from the $UE$, as the distance between the reflectors increases, the accuracy of distinguishing between single reflector and double reflector case also increases and it reaches its maximum when the two reflectors are diagonally opposite which is the maximum distance between them for a particular distance of the reflectors from the $UE$. It can be also observed that, if the reflectors are very close (\textcolor{black}{$0.5m$}) to $UE$ or very far (\textcolor{black}{$10m$}) from $UE$, the performance accuracy decreases.

Figure \ref{accu}(b) shows the variation of average accuracy w.r.t. the intra reflector distance for a particular distance of the reflectors from the SDR based LTE $UE$. The distances of the reflectors from the $UE$ are fixed to be \textcolor{black}{$4m$} here. Six different intra reflector distances are considered as explained earlier. The absolute straight line distances between the reflectors are  \textcolor{black}{$0.00m$, $2.50m$, $3.06m$, $5.65m$, $7.39m$ and $8.00m$} respectively. It can be observed from the figure that as the distance between the reflectors increases, the accuracy increases. This means that the capability of LTE-CommSense to distinguish between two reflector case also increases.

In the next case, we keep the reflectors diagonally opposite and vary the distance ($D$) of them from the SDR based LTE $UE$. Data is collected for five different distances of the reflectors from the $UE$ viz. \textcolor{black}{$0.5m$, $2m$, $4m$, $7m$ and $10m$}. It can be observed from Figure \ref{accu}(c) that, if the reflectors are very close to $UE$ (\textcolor{black}{$0.5m$}), the accuracy is lower. This is due to the reason that, the absolute distance of the reflectors will be very less. Therefore, it is difficult for CommSense to distinguish between single and double reflectors. Up to \textcolor{black}{$7m$} distance, the accuracy is $\sim 90\%$ but beyond that the accuracy decreases gradually. This is mainly due to the fact that, LTE-CommSense utilizes communication infrastructure and the operating signal to noise ratio (SNR) for a communication system is lower than that of a radar system. For LTE communication system, typical SNR level is $\sim7dB$ for a fair signal quality whereas, radar operation at long ranges requires SNR of $\sim25dB$ \cite{snr_ref}. If higher accuracy is required beyond \textcolor{black}{$10m$} for LTE-CommSense instrument, then the operating SNR level has to be improved further. This demands modification in the communication infrastructure being used for CommSense which is against its basic principle. Additionally, as the distance between this passive instrument and the object of interest increases further, environmental effects viz. fading, noise, received signal strength, near range ground bouncing, multi-path reflections form nearer objects etc.) will dominate the received signal as can be seen from \cite{06373911}. This will decrease the accuracy gradually with respect to increase of distance. Increasing the size of the reflector should further increase the range over which the accuracy is more than $90\%$. 

A detection methodology should optimize both FAR and FRR. Figure \ref{accu}(d) shows the FAR and Figure \ref{accu}(e) the FRR values with respect to the distances of reflector from the SDR based LTE $UE$. It can be observed that, FAR are very less and FRR are very high when the distance between them are less because, in that type of situation. it is difficult for the CommSense device to distinguish between single and double reflector case. Gradually FAR increases up to a maximum of $0.16$ for \textcolor{black}{$7m$} distance between the reflectors and the $UE$. But as the distance between the $UE$ and the reflectors increases further, e.g. for \textcolor{black}{$10m$} distance, FAR decreases. This is due to the fact that, as distance increases, less amount of DL signals are reflected back to the $UE$ by the reflectors. We also observe that, as the distance between them increases, FRR gradually decreases.

\subsection{Derivation of Resolution according to Neyman Pearson Criteria}

Using Neyman-Pearson principle, resolution $(\delta_{NP})$ is the maximum change in the `artifact of interest', $\Delta \theta$, to ascertain that the change in the false present rate $\Delta FAR \leq $ an arbitrary small number. We consider the change in the distance between the reflectors as the `artifact of interest'. From the figure on FAR for this experiment (Figure \ref{accu}(d)), changes in the intra reflector distances and corresponding FAR changes for a particular distance of the reflectors from the $UE$ can be recorded. Cases of no FAR value changes corresponding to maximum distance change between the reflectors for different distances from the $UE$ can be evaluated from that. We observed that, for our proposed instrument, when the distance from $UE$ is \textcolor{black}{$0.5m$} and the distance of the reflectors are $0.3827 \approx 0.38m$ along the circumference of the circle with $UE$ in center and \textcolor{black}{$0.5m$} as radius, the change in the value of FAR is arbitrary small. Also, when the distance between the reflector increases, FAR also increases. Therefore, it can be concluded that, \textcolor{black}{$0.3827 \approx 0.38m$} is the minimum distance change for which the change in FAR value is smaller than any arbitrary small number when the reflectors are \textcolor{black}{$0.5m$} apart from the $UE$. Following the same logic, the resolution of LTE-CommSense at distances \textcolor{black}{$2m$, $4m$, $7m$ and $10m$} can be evaluated as \textcolor{black}{$1.5m$, $2.5m$, $3m$ and $3.2m$} respectively.

\subsection{Derivation of Resolution according to Cramer Rao Lower Bound Principle}

The derivation of Cramer Rao lower bound principle can be found in (\cite{trees,kay}). Following `Definition 2' in Section \ref{crsec}, Cramer-Rao principle based resolution is calculated for each value of reflectors distance from user equipment ($UE$). To perform that we need the probability density function which is estimated as follows. Therefore, total six readings were taken for a particular distance ($D$) of $UE$ from the reflectors. Similarly five different $D$ values were considered. Additionally one reading were taken for the case of no reflectors were present. Hence, a total of thirty one readings were taken. One LTE frame is of ten milliseconds (ms) duration consisting ten subframes of one millisecond each containing I-Q data. If one recording of LTE DL data is captured for fifteen seconds, we can evaluate up to 1500 different channel estimates forming 1500 readings from a single recording as per the above procedure. In practice, downlink data were captured more than 15 seconds because a part of the initial data is consumed for timing and frequency offset correction operation. The procedure of timing and frequency offset correction is detailed in the LTE standards (\cite{lte_spec,lte1,lte2}). Therefore, we have 1500 readings for each of the intra reflector distances for each of the thirty one set of captured downlink data.

The accuracy values for each of the above cases are evaluated. For all the cases of distances of the reflectors from the $UE$ and intra reflector distances, the accuracy values assumes a Gaussian distribution with different mean and variance values. From the practical data, the mean, variance and standard deviation of the data are calculated for each case.

For a particular distance of the reflectors from the $UE$, the steps performed to evaluate the corresponding resolution are summarized in the pseudo-code shown in Algorithm \ref{crlb_pcode}. Here the number of different reflector distances from the $UE$ is denoted by $M$. Also, for every particular reflector distance from $UE$, the number of considered intra-reflector distances are denoted by $N$.

\begin{algorithm}
    \SetKwInOut{Input}{Input}
    \SetKwInOut{Output}{Output}
    \underline{function: CRLB-based-Resolution} ($\theta$, $P$, $Resolution$)\;
    \Input{$\theta$: represents intra-reflector distances for different reflector distances from $UE$, Evaluated $pdf$ ($P$) for multiple Reflectors Configurations}
    \Output{$Resolution$: Resolution of LTE-CommSense at different distances from the LTE $UE$}
\For{($i=0,i<M;i++$)}
{
\For{($j=0,j<N;j++$)}
{
	{\textbf DO} Derivation of $pdf$ (Curve Fitting)\\
	\eIf{$pdf$ is Gaussian} 
	{
			$P(\theta_{ij}) = \frac{1}{\sqrt{2\pi}\sigma}e^{-\frac{(\theta_{ij}-\mu_{ij})^2}{2\sigma_{ij}^2}} $
	}
	{
			Derive parameters for fitted $pdf$ \\
	}
    \eIf{($P(\theta_{ij}) = \frac{1}{\sqrt{2\pi}\sigma}e^{-\frac{(\theta_{ij}-\mu_{ij})^2}{2\sigma_{ij}^2}} $)} 
      {
$L(\theta_{ij}) = log_e(P(\theta_{ij})) = log_e(\frac{1}{\sqrt{2\pi}\sigma_{ij}}) -\frac{(\theta_{ij}-\mu_{ij})^2}{2\sigma_{ij}^2}$; \% log-likelihoods \\
$\frac{\partial L(\theta_{ij})}{\partial \theta_{ij}} = \frac{\mu_{ij}-\theta_{ij}}{2\sigma_{ij}^2} $; \% partial derivatives\\
{\textbf Evaluate:} $\biggr[\frac{\partial L(\zeta;\theta_{ij})}{\partial \theta_{ij}}\biggr]^2$;\% square of derivatives \\
$I(\theta_{ij}) = E\biggr[\biggr(\frac{\partial L(\zeta;\theta_{ij})}{\partial \theta_{ij}}\biggr)^2\biggr]$; \% information measure 
\\
$Var_{ij}=\frac{1}{I(\theta_{ij})}$; \% variances of the estimate\\
      }
      {
        If $pdf$ not Gaussian, derive above steps for that $pdf$.
      }
}
\For{($j=0,j<N;j++$)}
{
{\textbf Evaluate:} $\Delta \theta_{ij}= \theta_{ij}(M) - \theta_{ij} (M-1)$ ;\% Artifact of Interest\\
{\textbf Evaluate:} $\Delta Var_{ij} = VAR(\Delta \theta_{ij})$;\\
    \eIf{($ \Delta Var_{ij} <Var_{min}(i)  $)}
{
        $Var_{min}(i) = \Delta Var_{ij}$;\\
}
{
	$Var_{min}(i) = Var_{min}(i)$;\\
}
\For{$Var_{min}(i)$}
{
Find corresponding $\Delta \theta_{i}$;\\
$Resolution(i) = \Delta \theta_{i}$; \% Resolution at a Particular distance from $UE$\\
}
}
}
return $Resolution$\;
    \caption{Evaluation of proposed Cramer Rao Lower Bound (CRLB) based `Resolution' of LTE-CommSense from practical data for multiple object detection in outdoor environment}
    \label{crlb_pcode}
\end{algorithm}

The logarithm values ($L(\theta)$) obtained from practical data for different intra reflector distances for a particular distance of the reflectors from the $UE$, the derivative of $L(\theta)$ obtained for different intra reflector distances for a particular distance of the reflectors from the $UE$, $\biggr[\frac{\partial l(\zeta;\theta)}{\partial \theta}\biggr]^2$ values, expected value of $\biggr[\frac{\partial l(\zeta;\theta)}{\partial \theta}\biggr]^2$, Fisher information and variances are calculated using these data.

\begin{figure*}
\centering
\subfigure[]{\psfig{file=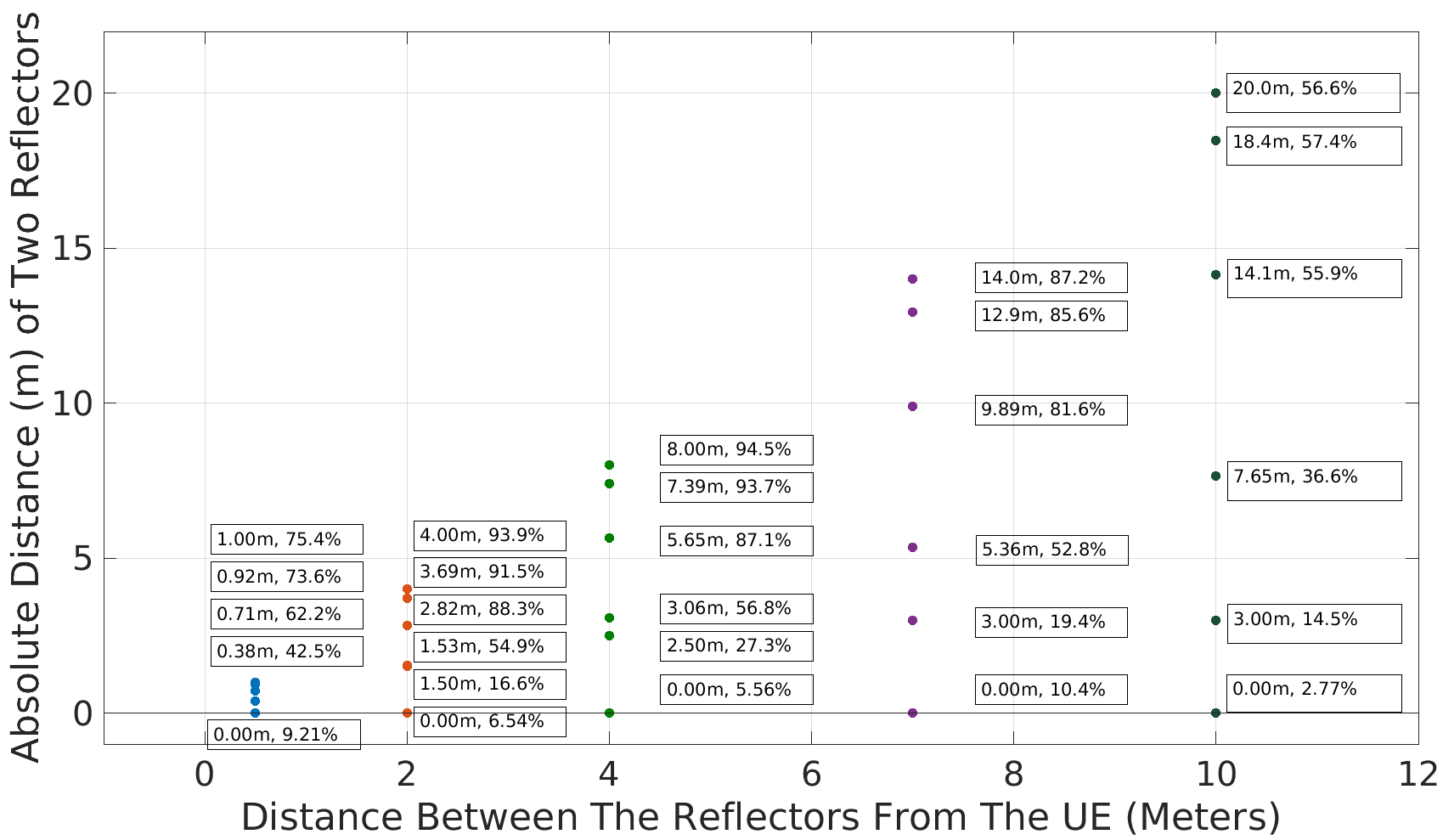,width=0.33\textwidth}}  
\subfigure[]{\psfig{file=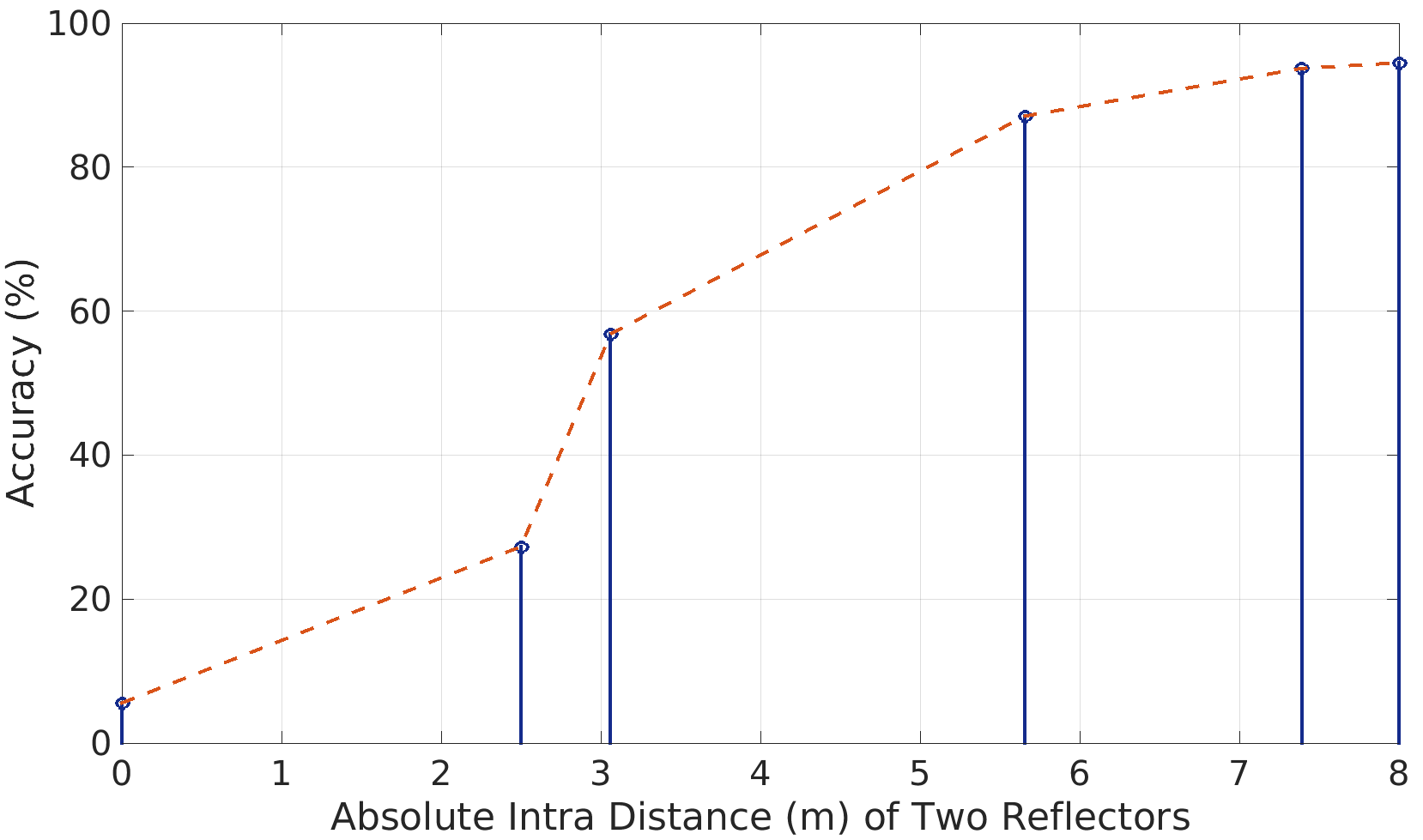,width=0.33\textwidth}}   
\subfigure[]{\psfig{file=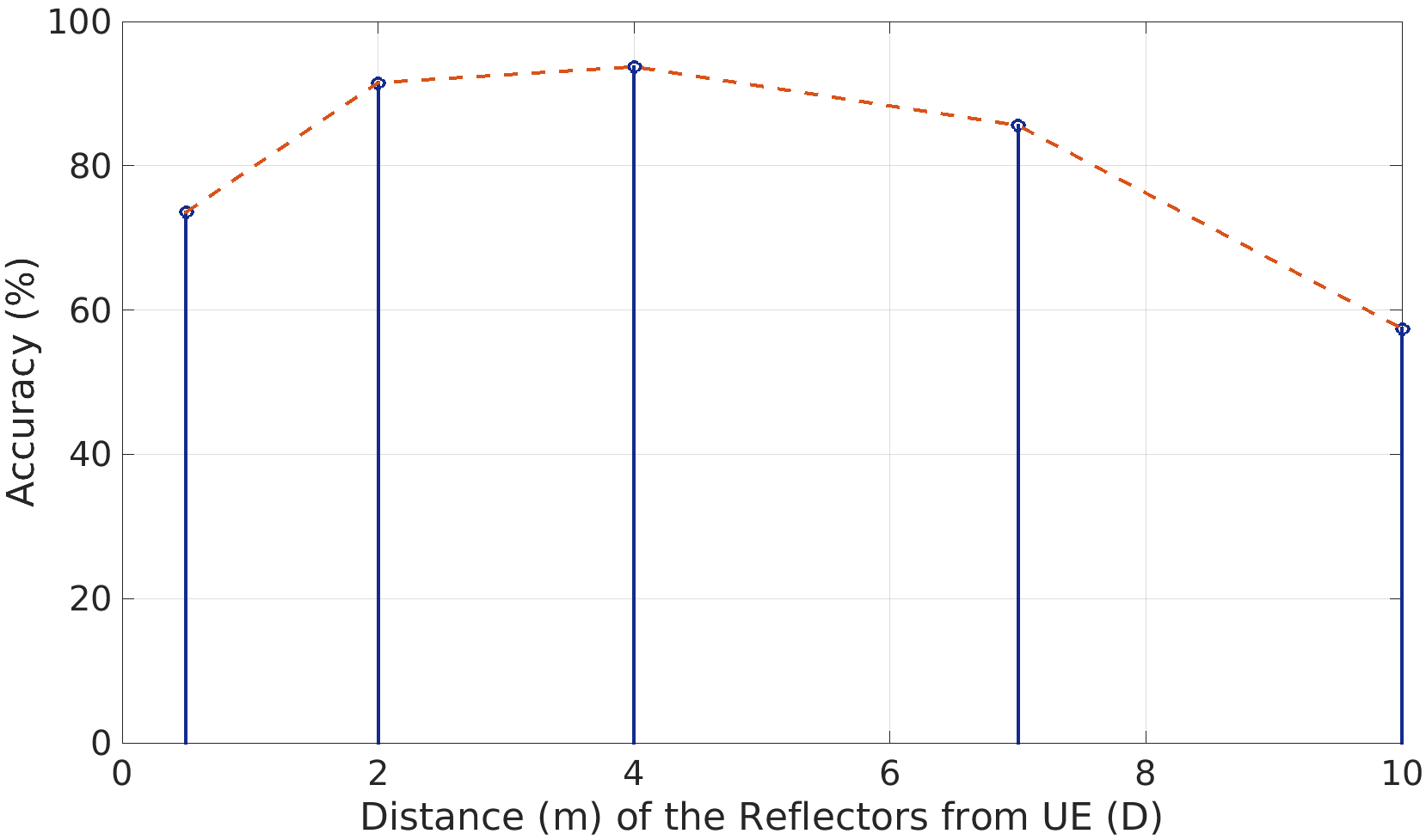,width=0.33\textwidth}}  
\subfigure[]{\psfig{file=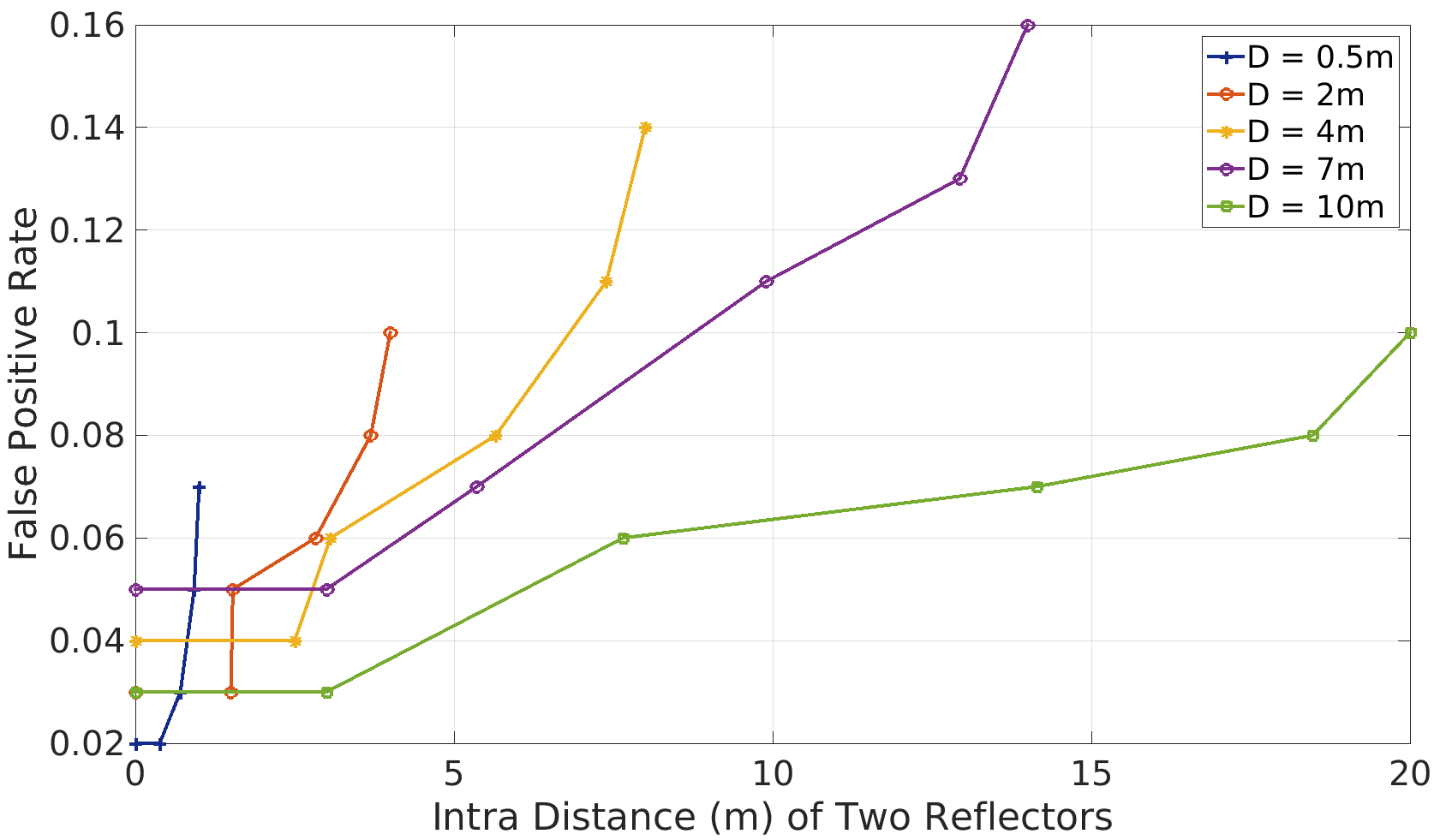,width=0.33\textwidth}}   
\subfigure[]{\psfig{file=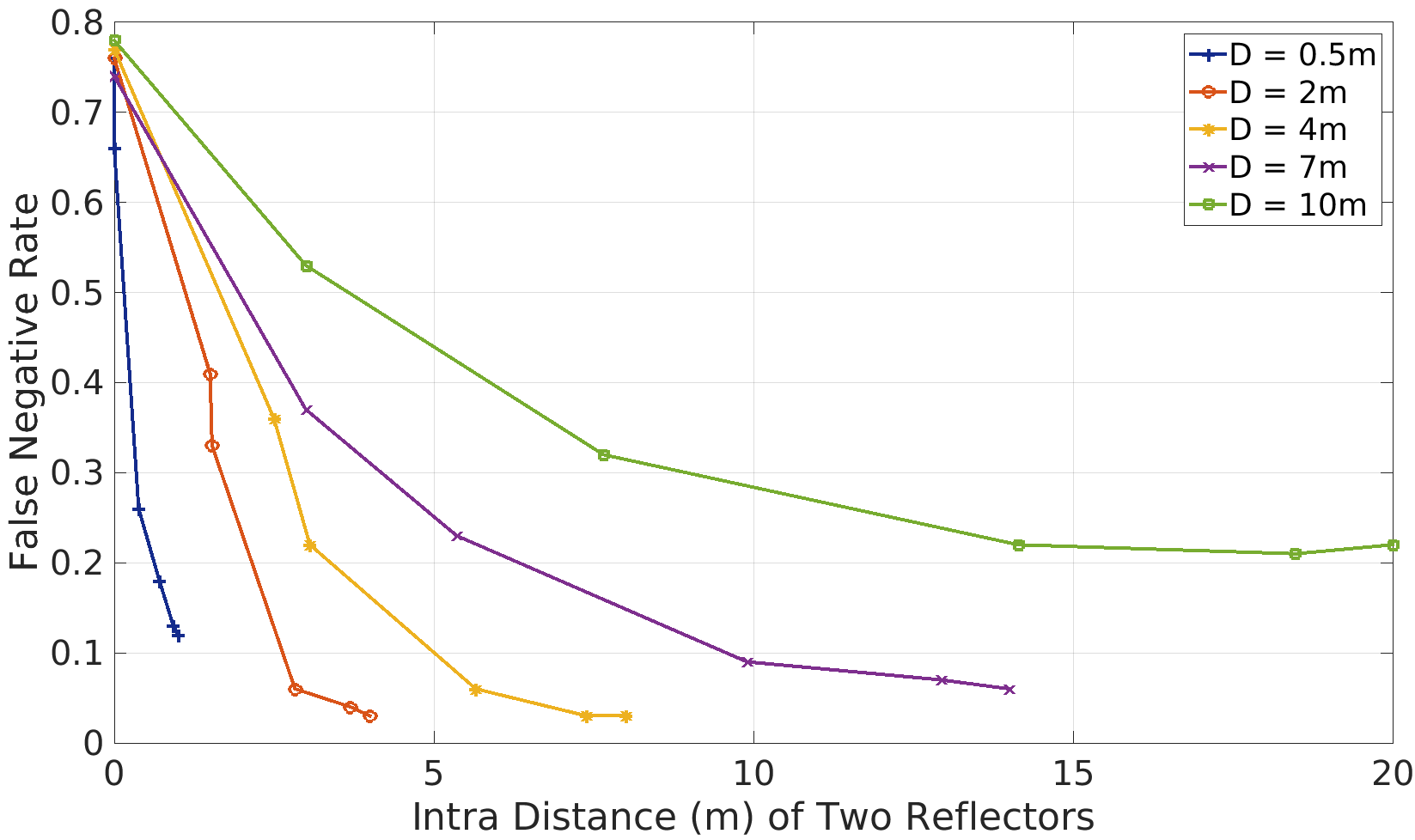,width=0.33\textwidth}}  
\subfigure[]{\psfig{file= 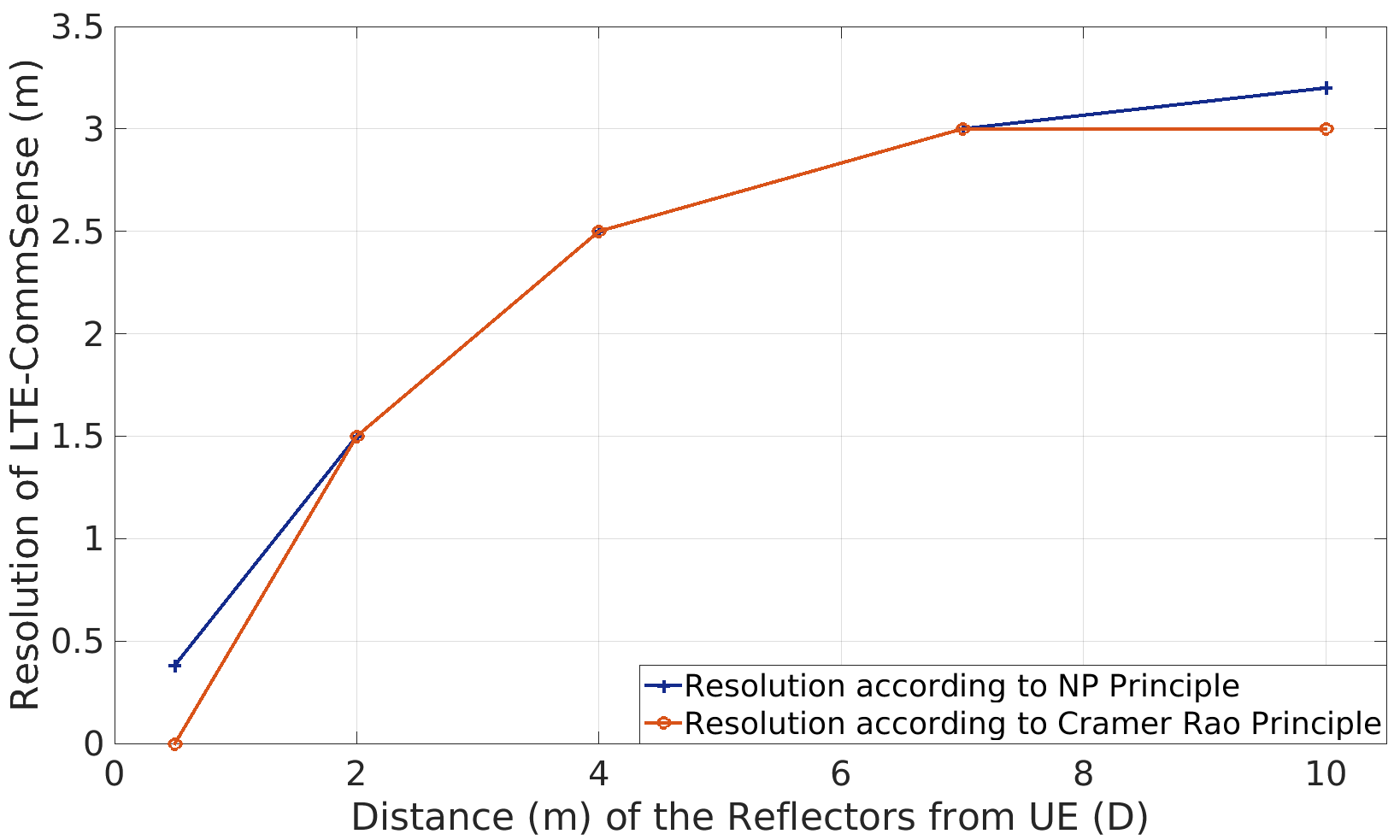,width=0.33\textwidth}}   
\subfigure[]{\psfig{file= 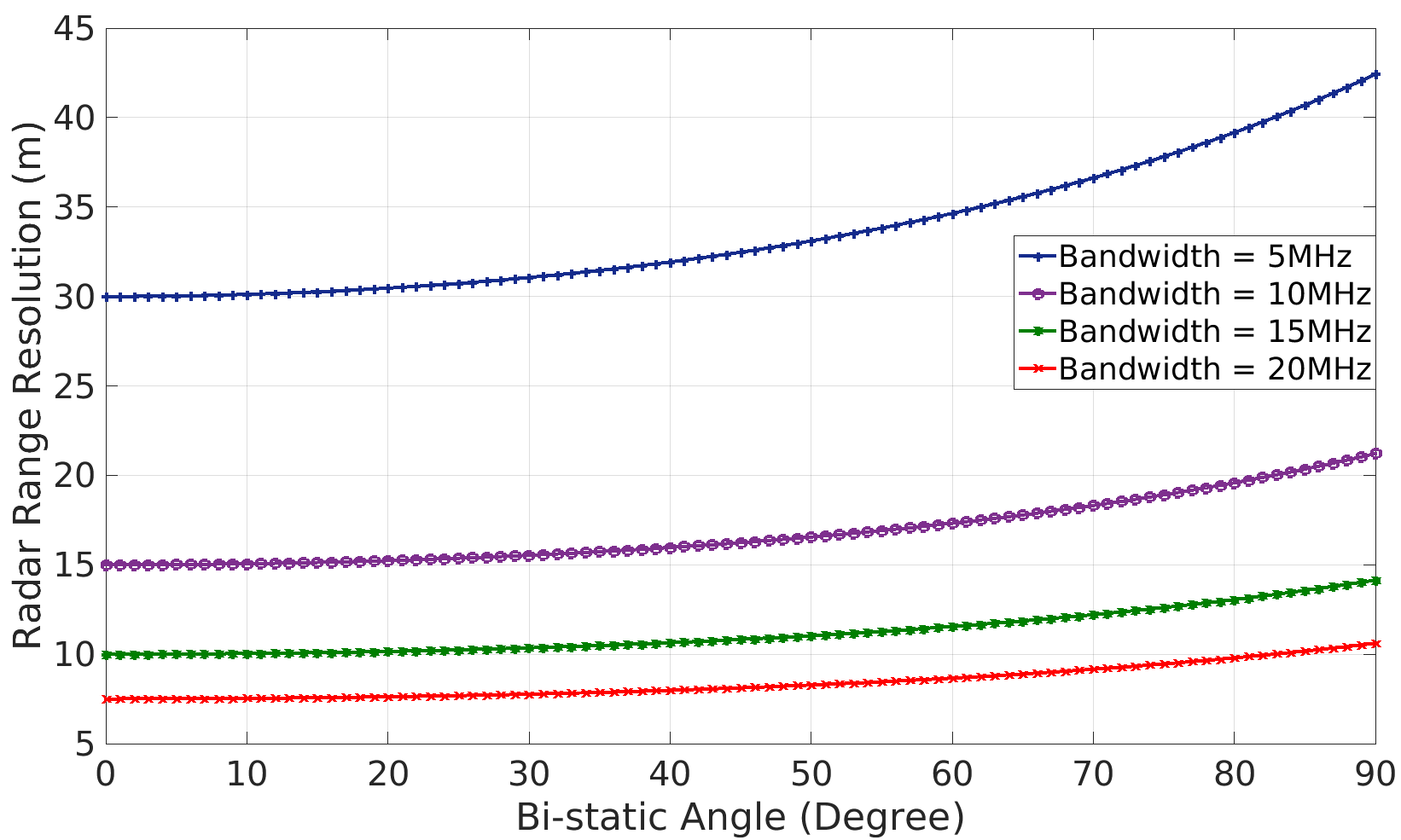,width=0.33\textwidth}}
\caption{\textcolor{black}{(a) Average Accuracies for Intra Reflector Distances in terms of Absolute Distance (m) from Each Other (in $Y$ axis) for different distances (m) of the reflectors from the $UE$ (in $X$ axis), (b) Variation of Average Accuracy w.r.t. Distance of the Reflectors from Each Other. For this figure, the Distance of both the Reflectors from the $UE$ is kept $04m$. Therefore, using Equation \ref{circ_dist} for our experimental setup, the considered distances ($S$) between the reflectors along the circumference are $0$, $1.2564m$, $3.141m$, $6.282m$, $9.423m$ and $15.564m$. Using these values of $S$ in Equation \ref{intra_abs}, the straight line distances between the reflectors are $0m$, $2.5m$, $3.06m$, $5.65m$, $7.39m$ and $8m$ respectively, (c) Variation of Average Accuracy w.r.t. Distance of the Reflectors from the $UE$. The Distance between the Reflectors is kept to be $2*D$, where $D$ is the distance of the reflectors from the $UE$, (d) FAR vs. Intra Reflector Distances in terms of Fraction of the Circumference for different distances of the reflectors from the $UE$ as Diameter, (e) FRR vs. Intra Reflector Distances in terms of Fraction of the Circumference for different distances of the reflectors from the $UE$ as Diameter, (f) Resolution of LTE-CommSense according to Neyman Pearson Principle and Cramer Rao Principle vs. the Distance of the Reflectors from the $UE$, (g) Bi-Static Radar Range Resolution (m) w.r.t. Bi-static Angle for different LTE Bandwidths. The bi-static angle is the angle exerted by the location of the target between the $eNodeB$ transmitter and $UE$.}}
\label{accu}
\end{figure*}

 The CRLB principle based derivation of resolution along with the NP principle \cite{kay} based resolution values for different distance of the reflectors from the $UE$ is shown in Figure \ref{accu}(f). As a comparison, we calculated the conventional range resolution of a bi-static radar system considering LTE $eNodeB$ as the transmitter the LTE $UE$ as the receiver. The available bandwidths in LTE (\cite{lte1,lte2}) are: $5$, $10$, $15$ and $20$ MHz \cite{lte_book}. In Figure \ref{accu}(g), we plotted the conventional bi-static radar range resolution (m) as a function of the bi-static angle for different bandwidths available for LTE communication system. The bi-static angle varies due to the different reflector positions as shown in Figure \ref{usrpn200}(b). 

The following observations can be inferred from the results:

\begin{enumerate} 
\item The derivation of resolution of LTE-CommSense for NP principle based method and CRLB principle based method yields similar results in this application scenario as is shown in Table \ref{res_table}.
\item It can be observed from Figure \ref{accu}(g) that the conventional radar range resolution depends on the radar operating bandwidths and the bi-static angle corresponding to bi-static configuration. For the bi-static radar scenario, the range resolution is constant when the bi-static angle and the bandwidth is constant. The best resolution is attained for the 20MHz bandwidth. Contrary to that, the CommSense principle based approach cannot provide range or angle information. The proposed resolution of LTE-CommSense is not a constant value. It varies with the distance of the reflectors from the $UE$.
\end{enumerate} 

\begin{table}[!h]
\caption{Derived Resolution of LTE-CommSense Instrument according to the two proposed definitions viz. Neyman Pearson based and Cramer Rao principle based definitions.}
\begin{center}
\begin{tabular}{|m{10mm}|m{28mm}|m{28mm}|}\hline
Distance (m) from UE & Resolution according to NP Principle (m) & Resolution according to CR Principle (m) \tabularnewline	\hline
0.5 &  0.3827  &	 0.0   	  \tabularnewline	\hline
2.0 &  1.5 		&	 1.5   	  \tabularnewline	\hline
4.0 &  2.5  	&	 2.5   	  \tabularnewline	\hline
7.0 &  3.0 		&	 3.0   	  \tabularnewline	\hline
10.0 & 3.2 		&	 3.0   	  \tabularnewline	\hline
\end{tabular}
\end{center}
\label{res_table}
\end{table}

\section{Conclusion}

LTE-CommSense is our proposed environment sensing system. It is a spectrum efficient system because it utilizes the communication spectrum without detrimental effect on it. As no transmitter and related regulations are involved, it can work undetected and may have smaller size and low cost compared to other environment sensing systems which require active transmitters. 

This work is an effort to apply the LTE-CommSense system for distinguishing objects in outdoor environment and evaluate the resolution of this system. As the objective of LTE-CommSense is to detect environment change due presence of objects or reflectors, we redefine resolution making `environment change' as the `artifact of interest'. We proposed two potential formulations of resolution for LTE-CommSense using Neyman Pearson principle and Cramer Rao lower bound (CRLB) principle. 

We have derived the performance parameters using practical data using the detailed experimental setup. LTE DL signal was captured using SDR based LTE receiver implementation with OpenLTE GNURadio framework and two corner reflectors. Multiple LTE DL signal was captured corresponding to different reflector distances from the $UE$ and different intra-reflector distances. LTE-CommSense detector analyzed these practical data and detector output is utilized to find out accuracy of this environment change detector, FAR and FRR of this system for all the experimental cases. These information were utilized further to evaluate the two proposed LTE-CommSense resolutions for the targeted application according to Neyman Pearson principle and CRLB principle. 

While calculating the resolution using Neyman Pearson principle, it was observed that, \textcolor{black}{$0.38m$, $1.5m$, $2.5m$, $3.0m$ and $3.2m$} are the value of the `artifact of interest' for which the FAR value is the smallest corresponding to $UE$ distances of \textcolor{black}{$0.5m$, $2m$, $4m$, $7m$ and $10m$} respectively. Therefore, the resolution of LTE-CommSense according to the NP principle is \textcolor{black}{$0.38m$, $1.5m$, $2.5m$, $3.0m$ and $3.2m$} corresponding to reflector distances of \textcolor{black}{$0.5m$, $2m$, $4m$, $7m$ and $10m$} from the $UE$ respectively. To calculate the resolution using Cramer Rao principle, we evaluated the variances of the estimates and the same was used to evaluate the resolution according to the proposed definition. It was found that, both the definitions yield similar values of resolution as expected as both the definitions signifies the same characterization parameter. We have compared the evaluated proposed resolution with the conventional radar range resolution for bi-static scenario. For a given bi-static angle and bandwidth, the conventional radar range resolution is constant. But the event dependent proposed resolution of LTE-CommSense detector depends on the distance of the reflectors from the user equipment.

Finally, we can successfully conclude that, LTE-CommSense can be used for object detection in outdoor environment. If detected, it can also distinguish between presence of one object and two objects. The exhaustive performance evaluation for this application using practical data may also serve as a benchmark and may be taken into consideration to apply LTE-CommSense for any other similar application.

\bibliographystyle{IEEEtran}
\bibliography{paper}

\begin{thebibliography}{10}
\providecommand{\url}[1]{#1}
\csname url@samestyle\endcsname
\providecommand{\newblock}{\relax}
\providecommand{\bibinfo}[2]{#2}
\providecommand{\BIBentrySTDinterwordspacing}{\spaceskip=0pt\relax}
\providecommand{\BIBentryALTinterwordstretchfactor}{4}
\providecommand{\BIBentryALTinterwordspacing}{\spaceskip=\fontdimen2\font plus
\BIBentryALTinterwordstretchfactor\fontdimen3\font minus
  \fontdimen4\font\relax}
\providecommand{\BIBforeignlanguage}[2]{{%
\expandafter\ifx\csname l@#1\endcsname\relax
\typeout{** WARNING: IEEEtran.bst: No hyphenation pattern has been}%
\typeout{** loaded for the language `#1'. Using the pattern for}%
\typeout{** the default language instead.}%
\else
\language=\csname l@#1\endcsname
\fi
#2}}
\providecommand{\BIBdecl}{\relax}
\BIBdecl

\bibitem{baker}
H.~Griffiths and C.~Baker, ``Passive coherent location radar systems. part 1:
  performance prediction,'' \emph{Radar, Sonar and Navigation, IEE
  Proceedings}, vol. 152, no.~3, pp. 153--159, 2005.

\bibitem{tong}
M.~Inggs and C.~Tong, ``Commensal radar using separated reference and
  surveillance channel configuration,'' \emph{Electronics letters}, vol.~48,
  no.~18, pp. 1158--1160, 2012.

\bibitem{Huang2015LowPO}
T.~Huang and T.~Zhao, ``Low pmepr {OFDM} radar waveform design using the
  iterative least squares algorithm,'' \emph{IEEE Signal Processing Letters},
  vol.~22, pp. 1975--1979, 2015.

\bibitem{7997194}
S.~D. Domenico, M.~D. Sanctis, E.~Cianca, P.~Colucci, and G.~Bianchi,
  ``{LTE}-based passive device-free crowd density estimation,'' in \emph{2017
  IEEE International Conference on Communications (ICC)}, May 2017, pp. 1--6.

\bibitem{199427}
G.~Papadopoulos, K.~Efstathiou, Y.~Li, and A.~Delis, ``Implementation of an
  intelligent instrument for passive recognition and two-dimensional location
  estimation of acoustic targets,'' \emph{IEEE Transactions on Instrumentation
  and Measurement}, vol.~41, no.~6, pp. 885--890, Dec 1992.

\bibitem{akm_patent}
A.~Mishra, ``Monitoring changes in an environment by means of communication
  devices.''\hskip 1em plus 0.5em minus 0.4em\relax Google Patents, Oct.~27
  2016, { WO} Patent App. PCT/IB2016/052,235.

\bibitem{bhatta}
A.~Bhatta and A.~K. Mishra, ``{GSM}-based {CommSense} system to measure and
  estimate environmental changes,'' \emph{IEEE Aerospace and Electronic Systems
  Magazine}, vol.~32, no.~2, pp. 54--67, February 2017.

\bibitem{phd_mag}
S.~Sardar, A.~K. Mishra, and M.~Z.~A. Khan, ``{LTE CommSense} for object
  detection in indoor environment,'' in \emph{IEEE Aerospace and Electronic
  Systems Magazine}, Accepted in July 2017.

\bibitem{7887698}
G.~Mois, S.~Folea, and T.~Sanislav, ``Analysis of three iot-based wireless
  sensors for environmental monitoring,'' \emph{IEEE Transactions on
  Instrumentation and Measurement}, vol.~66, no.~8, pp. 2056--2064, Aug 2017.

\bibitem{lte_book}
S.~Sesia, I.~Toufik, and M.~Baker, \emph{{LTE} - The {UMTS} Long Term
  Evolution, From Theory to Practice}.\hskip 1em plus 0.5em minus 0.4em\relax
  Wiley, 2009.

\bibitem{pheno}
J.~R. Huynen, ``A revisitation of the phenomenological approach with
  applications to radar target decomposition,'' \emph{DTIC Document, Tech.
  Rep.}, 1982.

\bibitem{7831367}
S.~Mühlbacher-Karrer, A.~H. Mosa, L.~Faller, M.~Ali, R.~Hamid, H.~Zangl, and
  K.~Kyamakya, ``A driver state detection system combining a capacitive hand
  detection sensor with physiological sensors,'' \emph{IEEE Transactions on
  Instrumentation and Measurement}, vol.~66, no.~4, pp. 624--636, April 2017.

\bibitem{8360051}
Z.~Khan, J.~J. Lehtomäki, E.~Hossain, M.~Latva-Aho, and A.~Marshall, ``An
  fpga-based implementation of a multifunction environment sensing device for
  shared access with rotating radars,'' \emph{IEEE Transactions on
  Instrumentation and Measurement}, vol.~67, no.~11, pp. 2561--2578, Nov 2018.

\bibitem{8349939}
H.~Zheng, R.~Ma, M.~Liu, and Z.~Zhu, ``A linear dynamic range receiver with
  timing discrimination for pulsed tof imaging ladar application,'' \emph{IEEE
  Transactions on Instrumentation and Measurement}, vol.~67, no.~11, pp.
  2684--2691, Nov 2018.

\bibitem{8326735}
H.~Lee and K.~Ke, ``Monitoring of large-area iot sensors using a lora wireless
  mesh network system: Design and evaluation,'' \emph{IEEE Transactions on
  Instrumentation and Measurement}, vol.~67, no.~9, pp. 2177--2187, Sept 2018.

\bibitem{6582678}
J.~Gutiérrez, J.~F. Villa-Medina, A.~Nieto-Garibay, and M.~Ã. Porta-Gándara,
  ``Automated irrigation system using a wireless sensor network and gprs
  module,'' \emph{IEEE Transactions on Instrumentation and Measurement},
  vol.~63, no.~1, pp. 166--176, Jan 2014.

\bibitem{7835628}
W.~Xue, W.~Qiu, X.~Hua, and K.~Yu, ``Improved wi-fi rssi measurement for indoor
  localization,'' \emph{IEEE Sensors Journal}, vol.~17, no.~7, pp. 2224--2230,
  April 2017.

\bibitem{7905996}
C.~Chen, Y.~Han, Y.~Chen, F.~Zhang, and K.~J.~R. Liu, ``Time-reversal indoor
  positioning with centimeter accuracy using multi-antenna wifi,'' in
  \emph{2016 IEEE Global Conference on Signal and Information Processing
  (GlobalSIP)}, Dec 2016, pp. 1022--1026.

\bibitem{7472878}
C.~Chen, Y.~Chen, H.~Lai, Y.~Han, and K.~J.~R. Liu, ``High accuracy indoor
  localization: A wifi-based approach,'' in \emph{2016 IEEE International
  Conference on Acoustics, Speech and Signal Processing (ICASSP)}, March 2016,
  pp. 6245--6249.

\bibitem{7489757}
A.~Bhatta and A.~K. Mishra, ``Implementation of {GSM} channel estimation using
  open-source {SDR} environment,'' in \emph{2015 International Conference on
  Microwave, Optical and Communication Engineering (ICMOCE)}, Dec 2015, pp.
  322--325.

\bibitem{africon}
S.~Sardar, A.~K. Mishra, and M.~Z.~A. Khan, ``{LTE-CommSense} system and its
  feasibility analysis,'' \emph{$13^{th}$ edition of IEEE AFRICON 2017}, 2017.

\bibitem{4071623}
G.~C. Gill and P.~L. Hexter, ``Some instrumentation definitions for use by
  meteorologists and engineers,'' \emph{IEEE Transactions on Geoscience
  Electronics}, vol.~11, no.~2, pp. 83--89, April 1973.

\bibitem{7882660}
Z.~Chen and W.~Li, ``Multisensor feature fusion for bearing fault diagnosis
  using sparse autoencoder and deep belief network,'' \emph{IEEE Transactions
  on Instrumentation and Measurement}, vol.~66, no.~7, pp. 1693--1702, July
  2017.

\bibitem{lte_spec_3}
{European Telecommunications Standards Institute (ETSI)}, ``{LTE}; evolved
  universal terrestrial radio access ({E-UTRA}); radio resource control
  ({RRC}); protocol specification,'' \emph{ETSI TS 136 331}, vol. 13.0.0, pp.
  1--670, 2016.

\bibitem{78300}
J.~D. Echard, ``Estimation of radar detection and false alarm probability,''
  \emph{IEEE Transactions on Aerospace and Electronic Systems}, vol.~27, no.~2,
  pp. 255--260, Mar 1991.

\bibitem{farfrr}
S.~Sardar, G.~Tewari, and K.~A. Babu, ``A hardware/software co-design model for
  face recognition using {C}ognimem neural network chip,'' in \emph{2011
  International Conference on Image Information Processing}, Nov 2011, pp.
  1--6.

\bibitem{Mishra16b}
A.~K. Mishra, ``{A}pplication {S}pecific {IN}strumentation {({ASIN}):} {A}
  bio-inspired paradigm to instrumentation using recognition before
  detection,'' \emph{CoRR}, vol. abs/1611.00228, 2016.

\bibitem{usrp}
S.~Yun and L.~Qiu, ``Supporting wifi and {LTE} co-existence,'' in \emph{2015
  IEEE Conference on Computer Communications (INFOCOM)}, April 2015, pp.
  810--818.

\bibitem{rfx2400}
\emph{{USRP} N200/N210 NETWORKED SERIES.
  $http://www.ettus.com/product/category/Daughterboards/RFX2400$}.

\bibitem{antenna}
\emph{{USRP} N200/N210 NETWORKED SERIES.
  $http://www.ettus.com/product/details/VERT2450$}.

\bibitem{openlte}
\BIBentryALTinterwordspacing
N.~{N}ikaein, R.~{K}nopp, F.~{K}altenberger, L.~{G}authier, C.~{B}onnet,
  D.~{N}ussbaum, and R.~{G}haddab, ``{D}emo: {O}pen{A}ir{I}nterface: an open
  {LTE} network in a {PC},'' in \emph{{MOBICOM} 2014, 20th {A}nnual
  {I}nternational {C}onference on {M}obile {C}omputing and {N}etworking,
  {S}eptember 7-11, 2014, {M}aui, {H}awai}, {M}aui, {UNITED} {STATES}, 09 2014.
  [Online]. Available: \url{http://www.eurecom.fr/publication/4371}
\BIBentrySTDinterwordspacing

\bibitem{lte_spec_2}
ETSI, ``{LTE}; evolved universal terrestrial radio access ({E-UTRA}); tdd home
  enode b ({HeNB}) radio frequency ({RF}) requirements analysis,'' \emph{ETSI
  TR 136 922}, vol. 9.0.0, pp. 1--78, 2010.

\bibitem{snr_ref}
S.~H. M.~A. Sadoon and B.~H. Elias, ``Radar theoretical study: Minimum
  detection range and maximum signal to noise ratio (snr) equation by using
  matlab simulation program,'' \emph{American Journal of Modern Physics},
  vol.~2, pp. 234--241, July 2013.

\bibitem{06373911}
P.~Krysik, P.~Samczynski, M.~Malanowski, L.~Maslikowski, and K.~Kulpa,
  ``Velocity measurement and traffic monitoring using a gsm passive radar
  demonstrator,'' \emph{Aerospace and Electronic Systems Magazine, IEEE},
  vol.~27, pp. 43--51, 10 2012.

\bibitem{trees}
H.~L.~V. Trees, \emph{Detection, {E}stimation, and {M}odulation {T}heory:
  {R}adar-{S}onar Signal Processing and {G}aussian Signals in Noise}.\hskip 1em
  plus 0.5em minus 0.4em\relax Melbourne, FL, USA: Krieger Publishing Co.,
  Inc., 1992.

\bibitem{kay}
S.~M. Kay, \emph{Fundamentals of Statistical Signal Processing: Estimation
  Theory}.\hskip 1em plus 0.5em minus 0.4em\relax Upper Saddle River, NJ, USA:
  Prentice-Hall, Inc., 1993.

\bibitem{lte_spec}
\emph{$http://www.3gpp.org/ftp/Specs/html-info/36-series.htm$}.

\bibitem{lte1}
D.~et. al., ``3g evolution: {HSPA} and {LTE} for mobile broadband,''
  \emph{Academic Press}, 2007.

\bibitem{lte2}
H.~E. et~al., ``Technical solutions for the {3G} long-term evolution,''
  \emph{IEEE Commun. Mag.}, vol.~44, no.~3, pp. 38--45, 2006.

\end{thebibliography}

\section{Appendix}

Figure \ref{lte_radar_time} represents a relative position of the LTE base station, $UE$ and a target inside the cell coverage area. As the objective is to calculate the best case timing precision calculation, the $UE$ is placed at the cell boundary. Let the height of the tower ($eNodeB$) be `$a$'; height of the person be `$b$'; distance between tower and person be `$c$' and the point where the ray bounces off from the target be `$bounce$'. Therefore, distance between the $eNodeB$ top and the target: 

\begin{equation}
d1 = \sqrt{a^2+bounce^2}
\end{equation}

Distance between the $UE$ and the target: 
\begin{equation}
d2 = \sqrt{b^2+(c-bounce)^2}
\end{equation}

Distance between the $eNodeB$ top and the $UE$: 
\begin{equation}
d3 = \sqrt{(a-b)^2+c^2}
\end{equation}

Therefore, distance difference between direct path and reflected path: 
\begin{equation}
\delta = d1+d2-d3
\end{equation}

We assume the speed of light is $3*10^8$ m/s. Therefore time difference between direct path and reflected path: 
\begin{equation}
\Delta t = \delta/(3*10^8 )
\end{equation}

In this work, we have assumed the height of the base station tower (`$a$') is \textcolor{black}{$10m$} and height of the $UE$ from the ground (`$b$') to be \textcolor{black}{$2m$}. The value of `$bounce$' is considered as half of the distance of the $UE$ from the LTE base station $eNodeB$.

\end{document}